\documentclass[aps,pra,showpacs,superscriptaddress,amssymb,twocolumn]{revtex4-1}
\usepackage{graphicx}
\usepackage{appendix} 
\usepackage{hyperref}
\usepackage{color}\hypersetup{colorlinks=true,citecolor=blue,linkcolor=black}
\usepackage{bm,amsmath}
\usepackage{dcolumn}
\input{Qcircuit}

\newcommand{\be}{\begin{equation}}
\newcommand{\ee}{\end{equation}}

\begin{document}

\rightline{\tt }

\vspace{0.2in}

\title{Measurement of GHZ and cluster state entanglement monotones in transmon qubits}

\author{Amara Katabarwa}
\email{akataba@uga.edu}
\author{Michael R. Geller}
\affiliation{Center for Simulational Physics, University of Georgia, Athens, Georgia 30602, USA}
\date{\today}

\begin{abstract}
Experimental detection of entanglement in superconducting qubits has been mostly limited, for more than two qubits, to witness-based and related approaches that can certify the presence of {\it some} entanglement, but not rigorously quantify how much. Here we measure the entanglement of three- and four-qubit GHZ and linear cluster states prepared on the 16-qubit IBM R\"ueschlikon (ibmqx5) chip, by estimating their entanglement monotones. GHZ and cluster states not only have wide application in quantum computing, but also have the convenient property of having similar state preparation circuits and fidelities, allowing for a meaningful comparison of their degree of entanglement. We also measure the decay of the monotones with time, and find in the GHZ case that they actually oscillate, which we interpret as a drift in the relative phase between the $|0\rangle^{\otimes n}$ and $|1\rangle^{\otimes n}$ components, but not an oscillation in the actual entanglement. After experimentally  correcting for this drift with virtual $Z$ rotations we find that the GHZ states appear to be considerably more robust than cluster states, exhibiting higher fidelity and entanglement at later times. Our results contribute to the quantification and understanding of the strength and robustness of multi-qubit entanglement in the noisy environment of a superconducting quantum computer.  
\end{abstract}

\maketitle

\section{INTRODUCTION}

Entanglement captures the intrinsic nonlocality of quantum systems and is a critical resource for quantum speedup. The crudest measure of entanglement is whether or not it is present: If the actual density matrix is a tensor product of single-qubit density matrices, it's not entangled. Witnesses  \cite{HorodeckiRMP09,GuhnePR09} have been used to establish entanglement in multi-qubit systems with up to 20 qubits \cite{NeeleyNat10,DiCarloNat10,SongPRL17,Friis171111092,Wang180103782}. When present, however, it is also interesting to quantify the degree of entanglement and its robustness to noise and decoherence \cite{NeeleyNat10,DiCarloNat10,13060510,150605353,180705572}. In this work we experimentally estimate entanglement monotones \cite{VidalJMO00,OsterlohPRA05,EltschkaPRA12} for the GHZ 
\begin{equation}
\frac{|0\rangle^{\otimes n} + |1\rangle^{\otimes n} }{\sqrt{2}}
\label{ghzDefinition}
\end{equation}
and linear (one-dimensional) cluster state
\begin{equation}
\prod_{i=1}^{n-1} {\rm CZ}_{i,i+1} |+\rangle^{\otimes n} 
\label{clusterDefinition}
\end{equation}
on the IBM  R\"ueschlikon (ibmqx5) chip \cite{urlnote}. Here ${\rm CZ}_{ii'}$ is the  gate ${\rm diag}(1,1,1,-1)$ on qubits $i$ and $i'$, and we have mapped the physical qubits to chains of length $n=3,4$. Specifically,  we measure the three-qubit monotone 
\begin{eqnarray}
\mathcal{E}_3 =  \frac{1}{3} \bigg\vert 
\langle XYY\mathcal{C} \rangle^2 
+ \langle ZYY\mathcal{C} \rangle^2 
-  \langle  IYY\mathcal{C}  \rangle^2 \nonumber  \\
+ \langle YXY\mathcal{C} \rangle^2 
+ \langle YZY\mathcal{C} \rangle^2 -  \langle  YIY \mathcal{C} \rangle^2 \nonumber \\
+ \langle YYX \mathcal{C} \rangle^2 + \langle YYZ \mathcal{C} \rangle^2 -  \langle  YYI \mathcal{C} \rangle^2 \bigg\vert ,
\label{E3def}
\end{eqnarray}
where $\mathcal{C}$ is the complex conjugation operator, and the four-qubit monotones
\begin{eqnarray}
\mathcal{E}_{4a} =  \langle YYYY \mathcal{C} \rangle^2  
\label{E4adef}
\end{eqnarray}
and
\begin{eqnarray}
\mathcal{E}_{4b} =   \bigg\vert 
\langle XYXY\mathcal{C} \rangle^2 + \langle ZYZY\mathcal{C} \rangle^2 -  \langle  XYIY \mathcal{C}  \rangle^2 \nonumber  \\
+ \langle ZYXY\mathcal{C} \rangle^2 + \langle ZYZY \mathcal{C} \rangle^2 -  \langle  ZYIY \mathcal{C} \rangle^2 \nonumber \\
- \langle IYXY \mathcal{C} \rangle^2 - \langle IYZY \mathcal{C} \rangle^2 + \langle  IYIY \mathcal{C} \rangle^2 \bigg\vert .
\label{E4bdef}
\end{eqnarray}
In this work we ignore the complex conjugation operator because our states are assumed to be real.

Entanglement monotones have several features that make them ideal for quantifying entanglement. First, they are strictly non-increasing under local (single-qubit) CPTP maps, making them less sensitive to decoherence than entanglement entropies and related measures based on reduced density matrices. Measuring their decay with time allows us to quantify and hopefully understand the environment of a noisy superconducting quantum computer and its effects on large-scale entanglement. Second, they allow for a direct comparison between different families of states and, if known, the maximum possible \cite{180711395}. The monotone $\mathcal{E}_3$, a symmetrized 3-tangle \cite{CoffmanPRA00}, is ideally equal to 1 (the maximum value) for both the GHZ and cluster states. (This is expected because the $n\!=\!3$ GHZ and linear cluster states are in the same entanglement class.) The monotone $\mathcal{E}_{4a}$ is the square of the 4-concurrence \cite{UhlmannPRA00,WongPRA01,OsterlohPRA17}, and is ideally equal to 1 for the GHZ state but vanishes for the cluster (showing that for $n\!=\!4$ they are in different classes). Therefore we cannot use  $\mathcal{E}_{4a}$ to quantify relative entanglement. We still measure it, however,  because it's a simple generalization of the squared two-qubit concurrence $ \langle YY \mathcal{C} \rangle^2$ \cite{WootersPRL98}. The monotone $\mathcal{E}_{4b}$ is ideally equal to 1 for both GHZ and cluster states, and vanishes not only on every four-qubit mixed product state 
\begin{equation}
\rho_1  \otimes \rho_2  \otimes \rho_3  \otimes  \rho_4 
\end{equation}
but also on partially entangled states of the form
\begin{equation}
\rho_{12} \otimes  \rho_{34},
\end{equation}
where $\rho_{12}$ and $\rho_{34}$ are entangled two-qubit states.
In other words, $\mathcal{E}_{4b}$ measures genuine four-qubit entanglement. The monotones $\mathcal{E}_{3}$ and $\mathcal{E}_{4a}$ also measure genuine multi-qubit entanglement in this same sense.

\begin{figure}
\includegraphics[width=7.5cm]{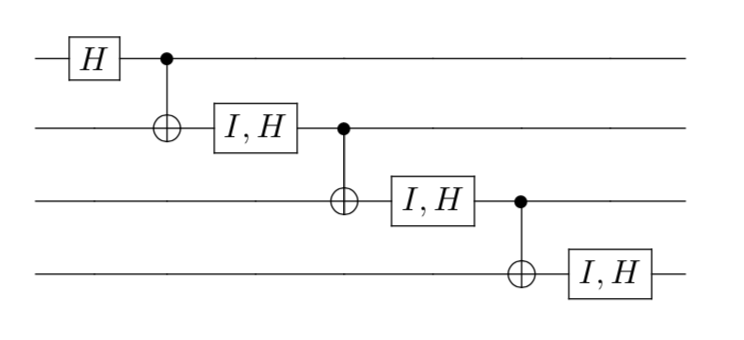} 
\caption{State preparation circuits for $n\!=\!4$. Single qubit gates 2 through $n$ are either identities ($I$) for GHZ or Hadamards ($H$) for the cluster state. The vertical gates are CNOTs.}
\label{statePrep}
\end{figure} 

A challenge of the monotone approach is that it relies on the antilinear conjugation operation $\mathcal{C}$. In this work we will attempt to bypass the complex conjugation step on the basis that the ideal GHZ and cluster states are real. This is an approximation that will limit the accuracy of our technique, and may also cause the monotones to have unphysical oscillations in time, unless the imaginary part is zeroed.  (We note that the approach of Di Candia {\it et al.} \cite{13060510} circumvents this limitation, at the expense of additional qubit and gate overhead.)

\begin{table}[htb]
\centering
\caption{State preparation error averaged over 32 independent circuit implementations, each estimated with 16 random readout-corrected Pauli expectation values. The preparation errors are written as the sample mean $\pm$ the standard error.}
\begin{tabular}{|c|c|c|c|c|}
\hline
Error (\%) & $n=3$ &  $n=4$ \\
\hline
GHZ & $11.83 \pm 0.51$ & $21.77 \pm 0.22$ \\
\hline
Cluster & $9.05 \pm 0.34$ & $21.17 \pm 0.29$ \\
\hline 
\end{tabular}
\label{StatePrepTable}
\end{table}

\section{STATE PREPARATION CIRCUITS AND FIDILITIES}

The state preparation circuits are shown in Fig.~\ref{statePrep}. The GHZ and cluster preparation circuits are the same except for $n-1$ Hadamards, and we confirm that their state preparation fidelities are very similar. This is important because it allows us to assume that the entangled states are prepared with similar fidelity.  To quantify this we prepare the $n=3,4$ GHZ and cluster states and measure their state preparation error (fidelity loss) by Flammia-Liu fidelity estimation \cite{FlammiaPRL11}. The results are summarized in Table \ref{StatePrepTable}. 

We find that the state preparation errors are quite noisy and (after readout correction) dominated by CNOT errors, which explains why the extra Hadamards in the cluster state preparation circuit do not, on average, result in a larger circuit error. In fact Table \ref{StatePrepTable} shows that the cluster states have slightly smaller state preparation errors. Histograms showing the distribution of state preparation errors are provided in Appendix \ref{STATE PREPARATION ERRORS}.

\section{MONOTONES}

\begin{figure}
\includegraphics[width=8cm]{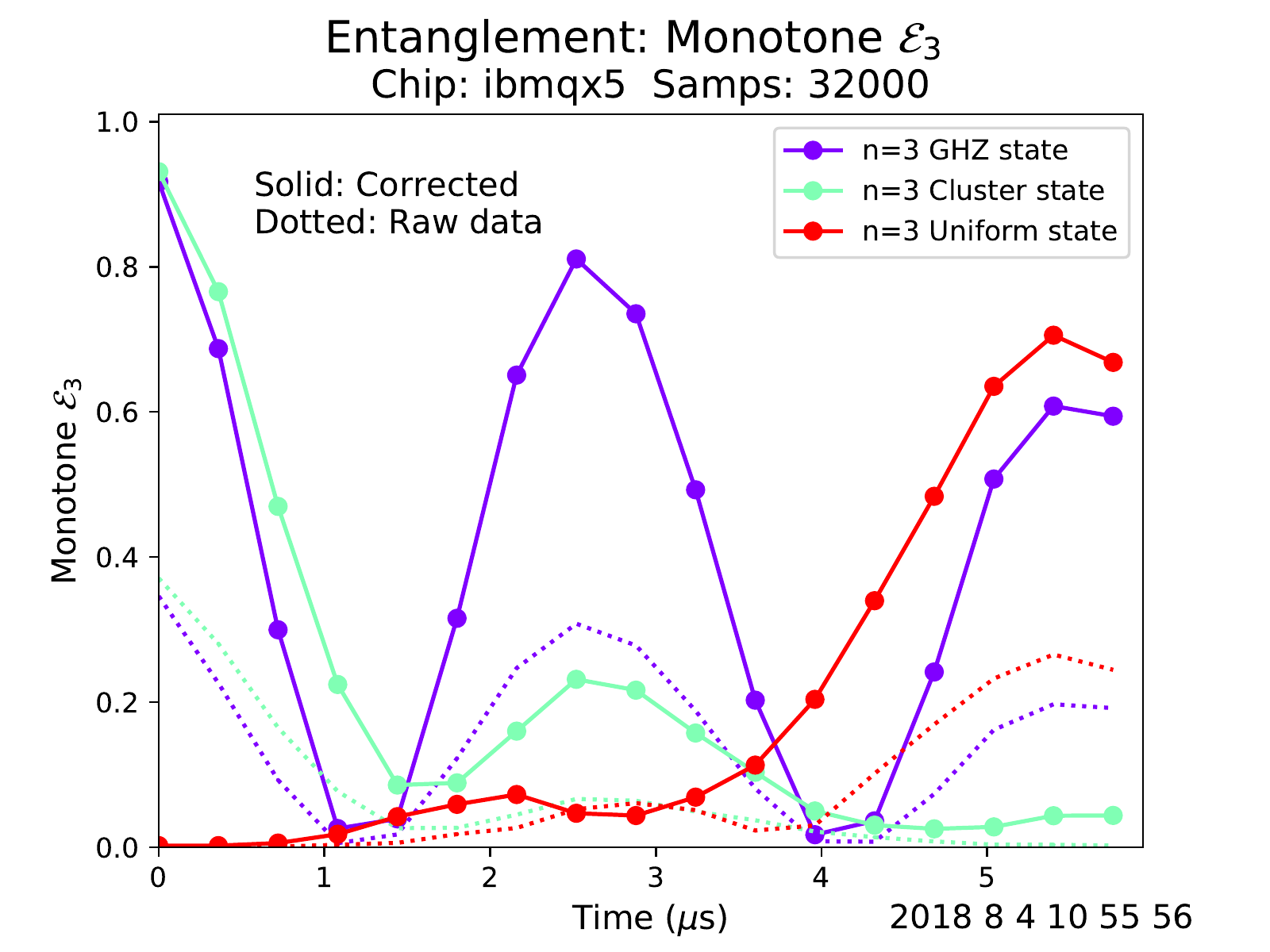} 
\caption{$\mathcal{E}_{3}$ versus time. The solid lines interpolate between readout-corrected measurements.}
\label{E3}
\end{figure} 

\begin{figure}
\includegraphics[width=8cm]{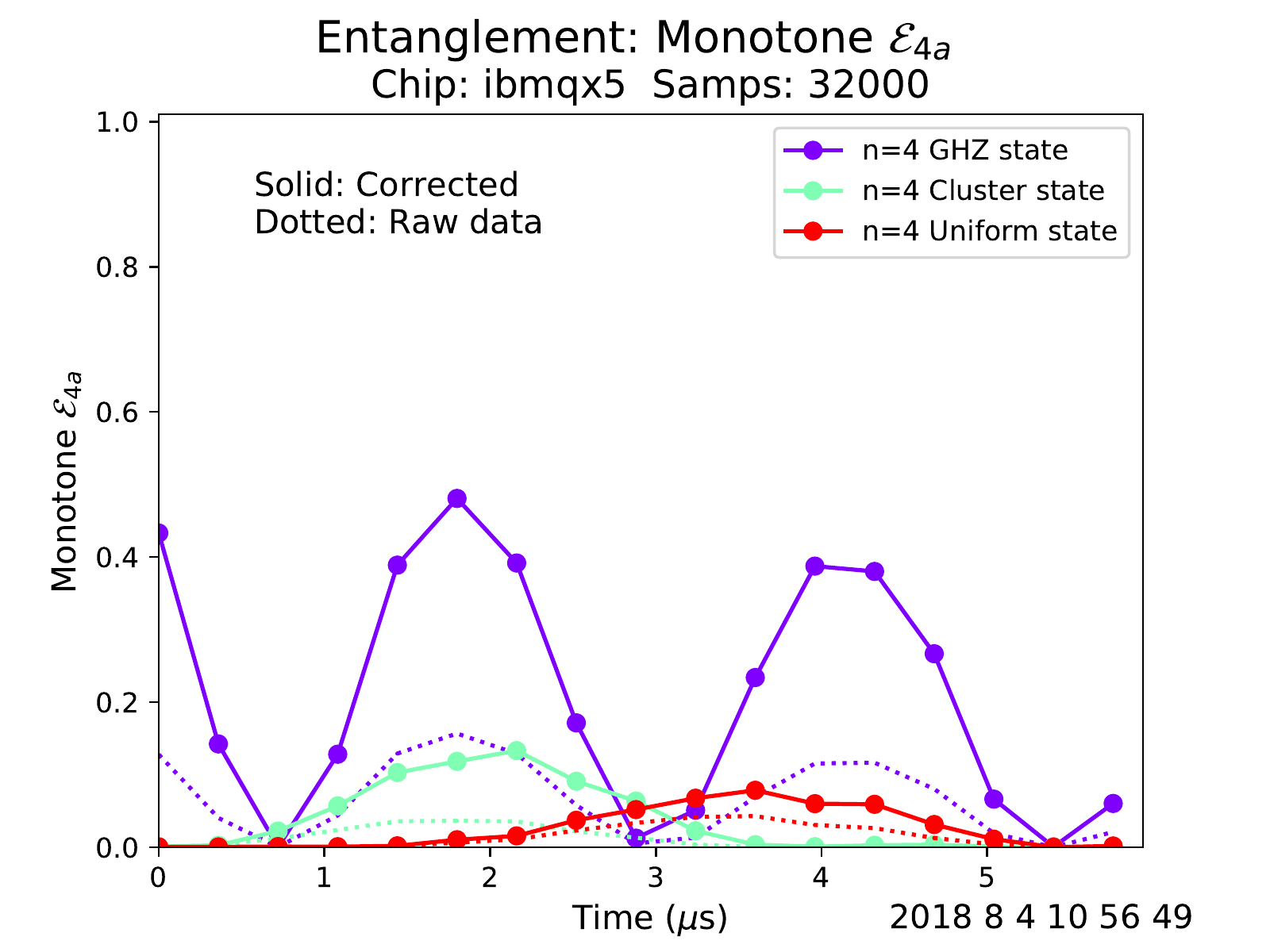} 
\caption{ $\mathcal{E}_{4a}$ versus time.}
\label{E4a}
\end{figure}

\begin{figure}
\includegraphics[width=8cm]{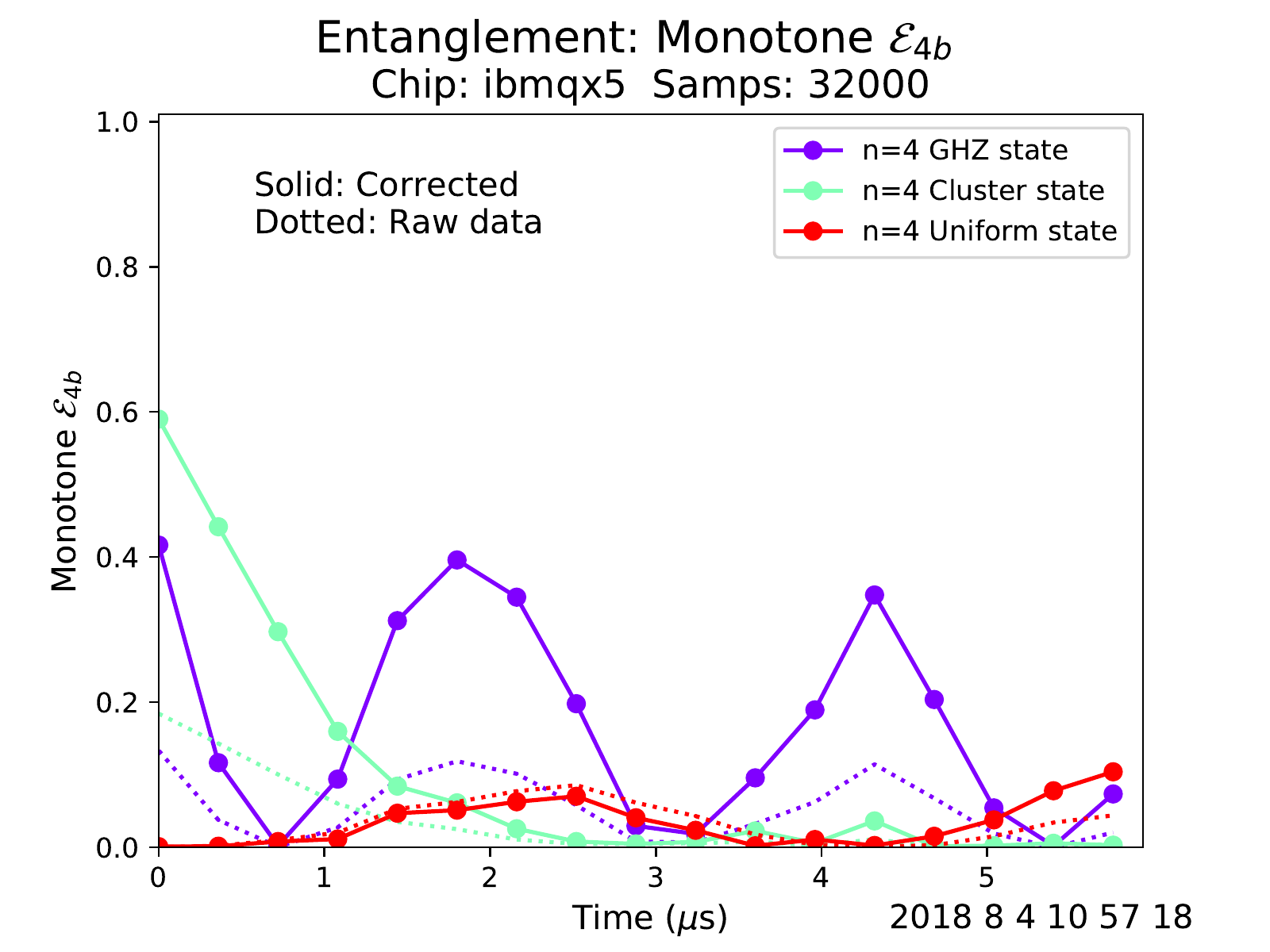} 
\caption{$\mathcal{E}_{4b}$ versus time.}
\label{E4b}
\end{figure}

\begin{figure}
\includegraphics[width=8cm]{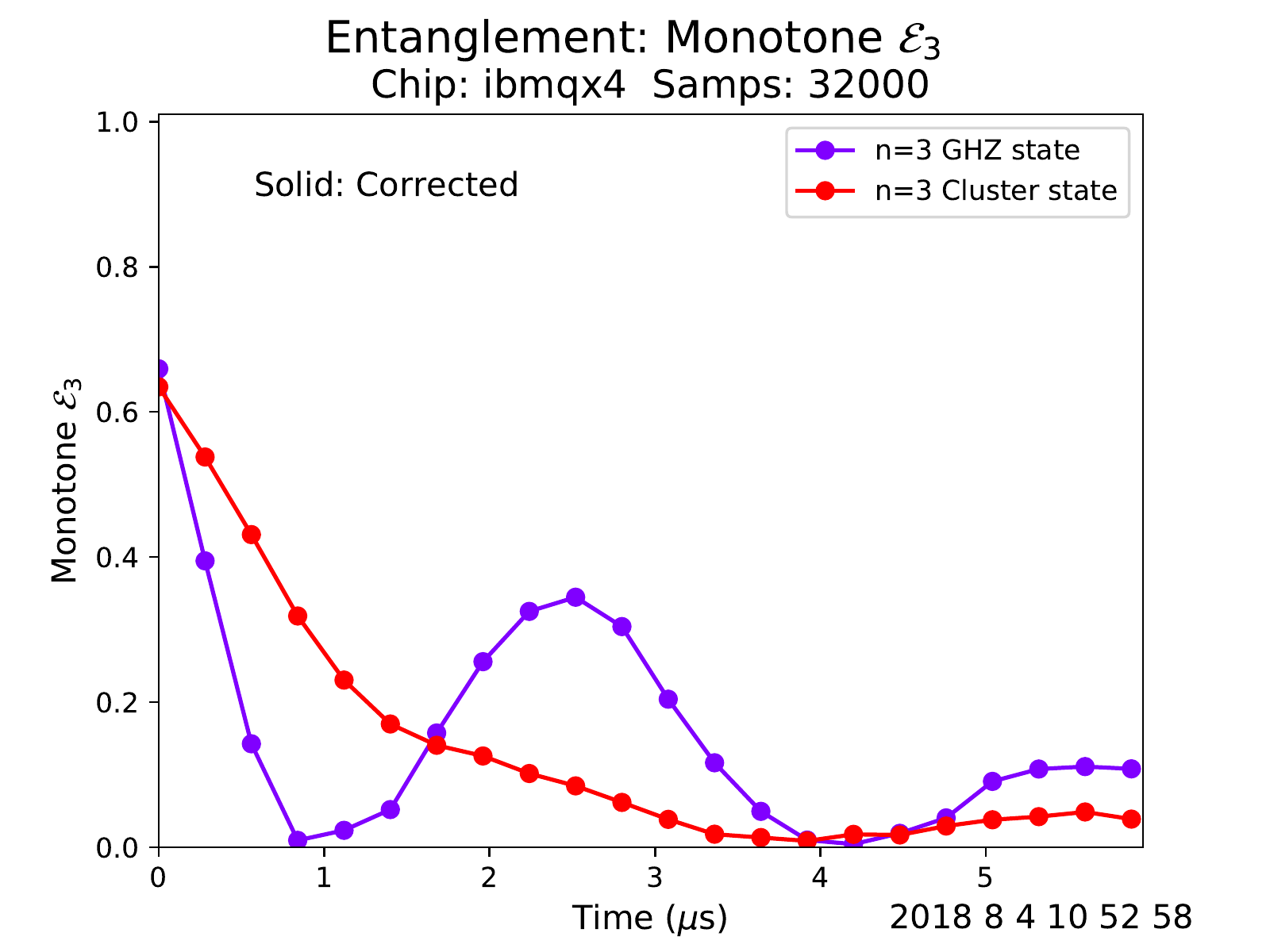} 
\caption{Oscillating $\mathcal{E}_{3}$ versus time on the IBM Tenerife (ibmqx4) chip.}
\label{E3_qx4}
\end{figure}

The measured monotone (\ref{E3def}) is shown in Fig.~\ref{E3}. Here $\mathcal{E}_3$ is measured after waiting for a time $t$. The delay is implemented by a sequence of 80ns identity gates. In addition to the GHZ and cluster states, we also measure the uniform state $ |+\rangle^{\otimes n}$ to include a nonentangled control subject. The observed oscillations in the data appear to violate the important non-increasing property of an entanglement monotone under local noise. The monotones (\ref{E4adef}) and (\ref{E4bdef}), shown in Figs.~\ref{E4a} and \ref{E4b}, are also non-monotonic. To confirm that the observed oscillations are not an artifact of the particular chip or qubits used, we also measured $\mathcal{E}_3$ on the 5-qubit IBM Tenerife (ibmqx4) chip,
and find oscillations with a similar frequency, as shown in Fig.~\ref{E3_qx4}.

\begin{figure}
\includegraphics[width=8cm]{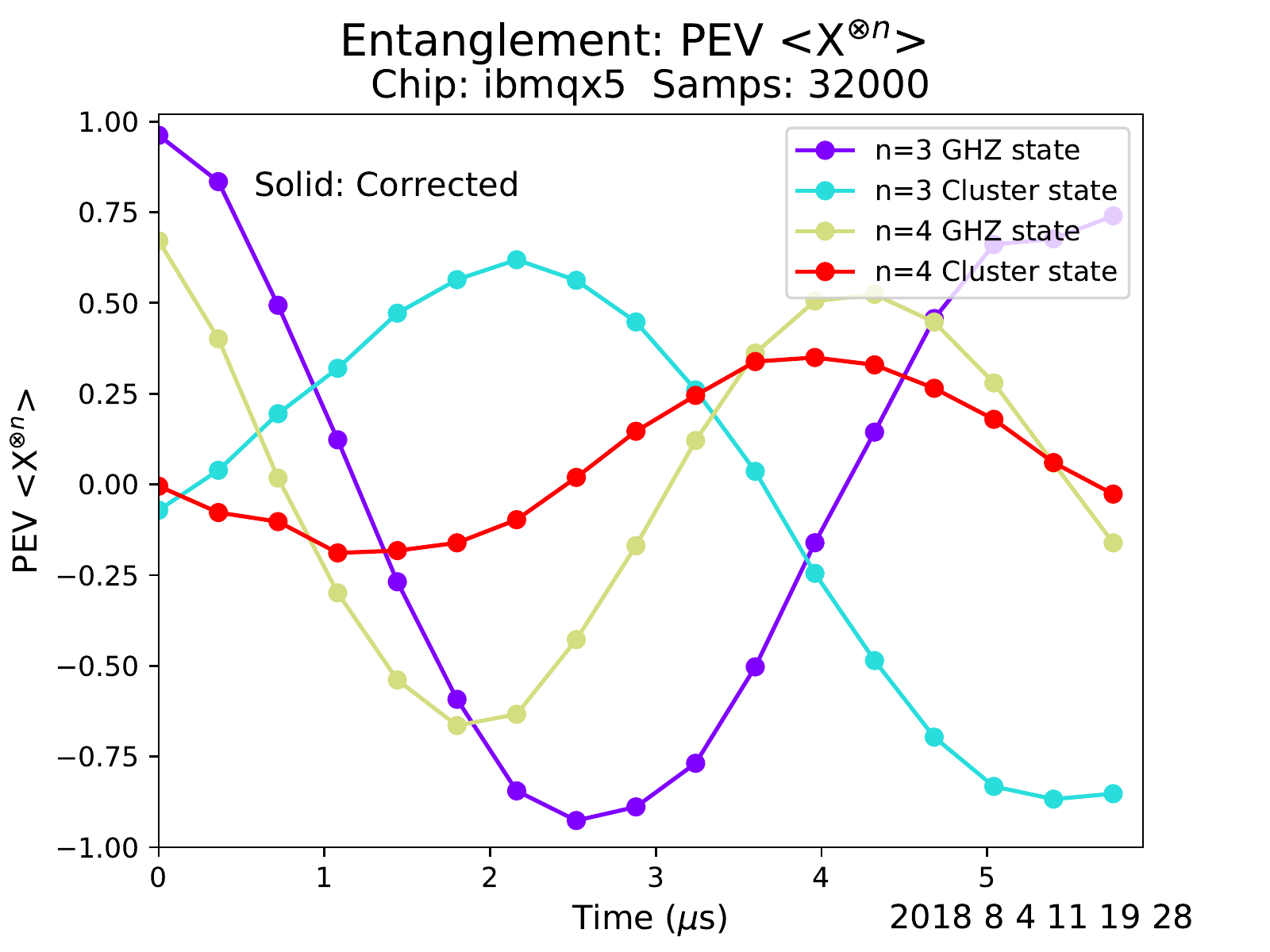} 
\caption{Pauli expectation values versus time.}
\label{X}
\end{figure}

Considering first the GHZ states, we interpret these oscillations as resulting from a nonzero relative phase $\phi$ 
between the $|0\rangle^{\otimes n}$ and $|1\rangle^{\otimes n}$ components, as defined in
\begin{equation}
\frac{|0\rangle^{\otimes n} + e^{i \phi} |1\rangle^{\otimes n} \! \! \! }{\sqrt{2}} .
\label{generalghzDefinition}
\end{equation}
Note that a nonzero $\phi \! \mod \pi$ results in an imaginary component in (\ref{generalghzDefinition}), invalidating our approach of neglecting the complex conjugations in the monotone definitions. To support this interpretation we measure the Pauli expectation value $ \langle X^{\otimes n} \rangle$, which can be used to extract the value of $\phi$. The data are shown in Fig.~\ref{X}. In the modified GHZ state (\ref{generalghzDefinition}),
\begin{equation}
\langle X^{\otimes n} \rangle  = \cos \phi.
\label{X^n}
\end{equation}
Fitting to the data we find that
\begin{equation}
\phi =  2 \pi  f_{\rm z} \,  t 
 \ \ {\rm with} \ \ 
f_{\rm z} \approx 
\begin{cases}
167 \, {\rm kHz} & {\rm for} \ \  $n=3$ \\
238 \, {\rm kHz} & {\rm for} \ \ $n=4$.
\end{cases}
\label{phi}
\end{equation}
The $n$ dependence of $f_z$, easily visible in Fig.~\ref{X}, suggests that each qubit has an unexpected 55 to 60 kHz splitting in the rotating frame. Interestingly, however, we did not observe oscillations in single-qubit Ramsey scans $\left( |0\rangle \rightarrow e^{i \pi Y/4} \ I_{{\rm delay} \,  t} \ e^{-i \pi Y/4} |0\rangle \right)$ versus $t$ on any of the individual qubits, but did in the concurrence $ \langle YY \mathcal{C} \rangle$ of a Bell state, suggesting that the phase drift (\ref{phi}) is intrinsically a multi-qubit effect.

Similar oscillations occur for the cluster state, but for this family the oscillation amplitude is much smaller and the oscillations are sometimes only barely visible before the entanglement has vanished. We do not have a simple model for the perturbed state in this case.

\section{Phase drift compensation}

\begin{figure}
\includegraphics[width=8cm]{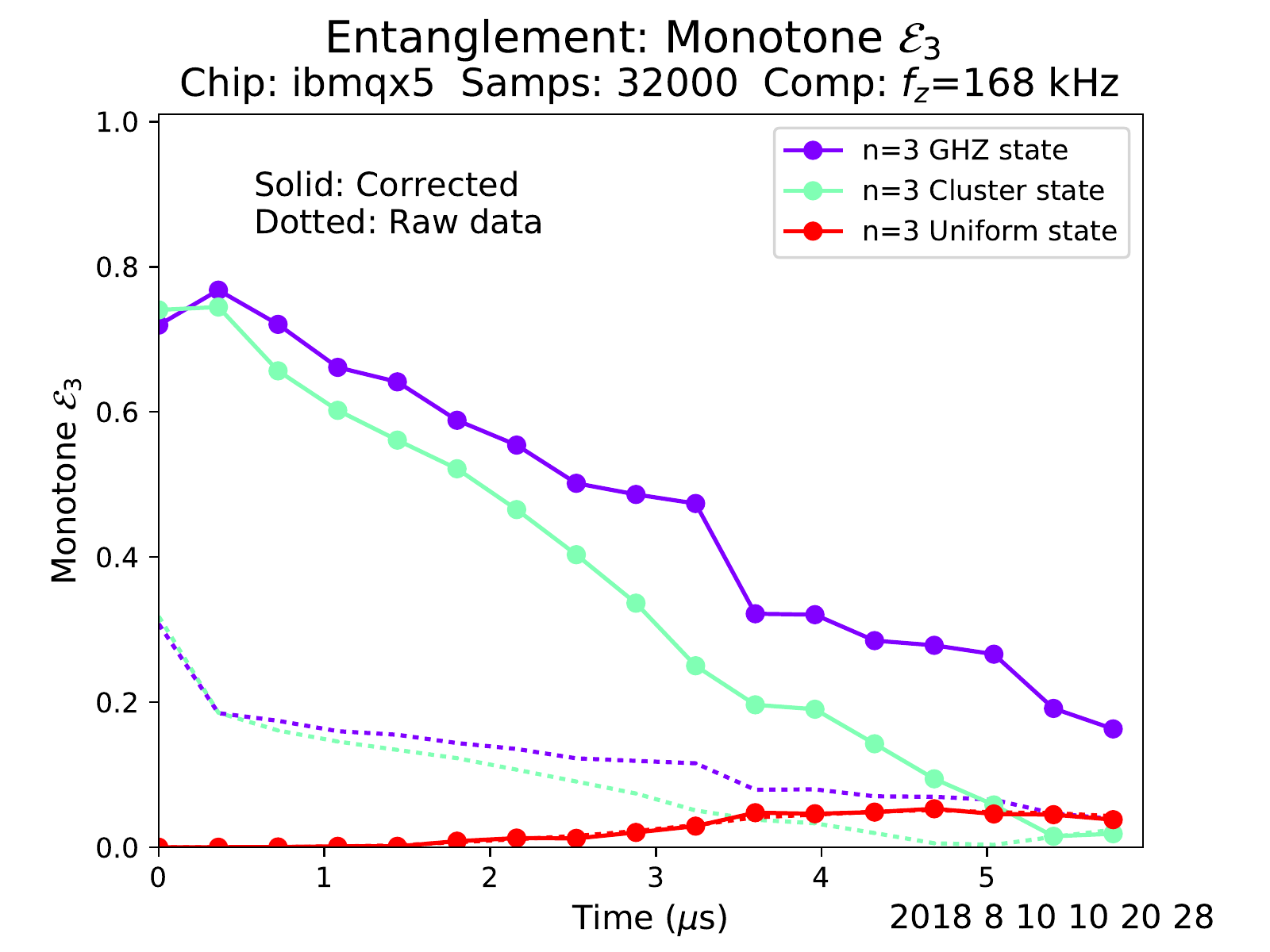} 
\caption{Same as Fig.~\ref{E3}, except that identical phase shifts of $-\phi/n$ are now applied to the qubits during the delay, to compensate for the phase drift. The GHZ state is clearly more robust than the cluster state at long times.}
\label{E3_comp}
\end{figure}

\begin{figure}
\includegraphics[width=8cm]{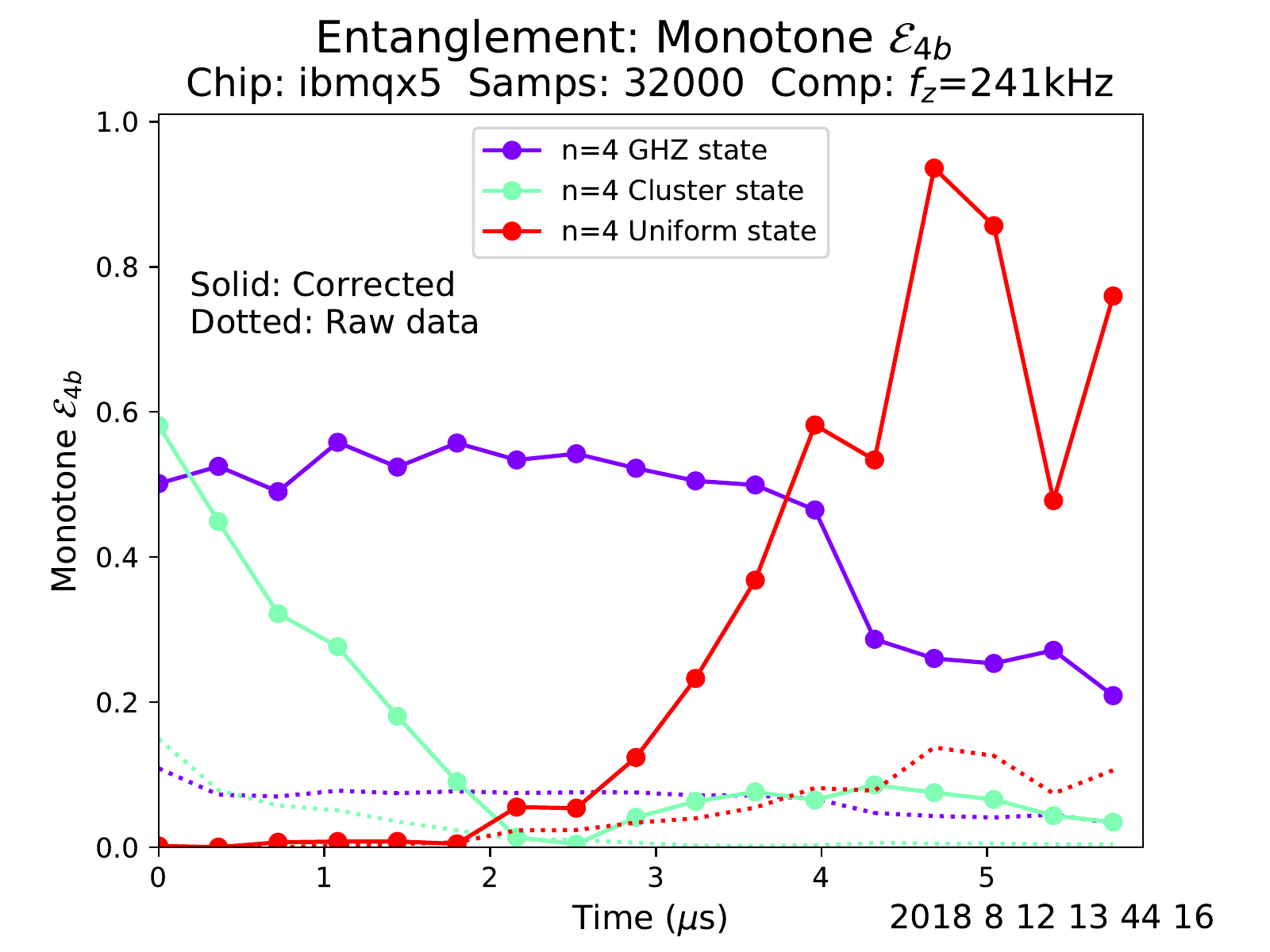} 
\caption{Same as Fig.~\ref{E4b}, but with phase compensation. Again the GHZ state appears to be much more robust.}
\label{E4b_comp}
\end{figure}

Having identified the phase drift (\ref{phi}) as the origin of the oscillations in the GHZ monotones, we attempt to cancel it by applying compensating virtual (software) $Z$ rotations with a phase opposite opposite to (\ref{phi}). Identical phase shifts of $-\phi/n$ are applied to each qubit. We apply this same correction to the GHZ, cluster, and uniform states. The results are shown in Figs.~\ref{E3_comp} and \ref{E4b_comp}. We refer to these corrected measurement results as {\it phase compensated} data.

The oscillations in $\mathcal{E}_{3}$ and $\mathcal{E}_{4b}$ are no longer present after phase compensation; they are now properly monotonic (apart from measurement noise). And after this correction, the GHZ states appear to be significantly more robust than the cluster states, which is the most interesting conclusion from this work.

The behavior of the uniform state $ |+\rangle^{\otimes n}$,  which we included as a (nominally) nonentangled control, is also interesting. In Fig.~\ref{E3}, the $n\!=\!3$ uniform state (red curve) appears to spontaneously develop a high degree of entanglement. However after phase compensation, Fig.~\ref{E3_comp}, entanglement is absent. This is consistent with our interpretation that the states in Fig.~\ref{E3} develop imaginary components,  invalidating our $\mathcal{E}_{3}$ measurement technique, which assumes purely real components. But the behavior of the $n\!=\!4$ uniform state is different: In Fig.~\ref{E4b}, the uniform state (red curve) shows no significant entanglement, but the phase compensated data does.

\section{CONCLUSIONS}

We have studied the relative strength and robustness of entanglement of the three- and four-qubit GHZ and linear cluster states on the IBM R\"ueschlikon (ibmqx5) superconducting quantum computer by measuring entanglement monotones $\mathcal{E}_{3}$ and $\mathcal{E}_{4b}$, defined in (\ref{E3def}) and (\ref{E4bdef}). These entanglement measures have the property that they are ideally equal to one for the GHZ and cluster states, allowing for a meaningful cross-comparison. However we find that $\mathcal{E}_{3}$ and $\mathcal{E}_{4b}$ are in fact non-monotonic, which we ascribe to the development of imaginary components of the states. After proposing a simple phase-drift model for this effect, we attempt to correct it, and after the correction find the GHZ states to be significantly more robust than the cluster states, exhibiting higher fidelity (data not shown here) and entanglement at later times. 

What is expected theoretically? There is a commonly stated  expectation that cluster states are more robust, due to their property that they remain partially entangled after measurement of a subset of $k<n$ qubits. (This property is essential for their use in measurement-based quantum computation.) We note, however, that this form of entanglement is not of the genuine multi-qubit type measured by $\mathcal{E}_{3}$ and $\mathcal{E}_{4b}$.
Furthermore, our results contradict the predictions of a Markovian $T_{1,2}$ model and a non-Markovian dephasing model with realistic (but spatially uncorrelated) $1/f$ flux noise. 

A simple explanation for the robustness of GHZ states observed here is that they only have two components (one relative phase) to acquire pure dephasing errors. And our compensation technique, which applies identical phase shifts $-\phi/n$ to each qubit, may be non-optimal for cluster states, whereas for GHZ states any distribution of phase shifts adding up to $-\phi$ is acceptable.
 
Finally, we note that the data presented here was acquired over more than 9 months, during which there were drifts in system parameters and gate fidelities, leading to small inconsistencies between some of the figures. However the principal observations, that GHZ and cluster state preparation fidelities and initial entanglement are similar, that $\mathcal{E}_{3}$ and $\mathcal{E}_{4b}$ oscillate in time, and that after phase compensation the GHZ states remain more entangled, were generally observed.
 
\begin{acknowledgments}

Data was taken on the IBM R\"ueschlikon (ibmqx5) chip, using the Quantum Experience API and the BQP software package developed by the authors. The complete data set represented here consists of $\sim$12k circuits, each measured 8000 times. We're grateful to the IBM Quantum Experience team for making their devices available. Thanks also to Matteo Mariantoni,  Phillip Stancil, Mingyu Sun, and Jason Terry for their  discussions and contributions to BQP. This work does not reflect the views or opinions of IBM or any of its employees. 

\end{acknowledgments}


\appendix

\section{STATE PREPARATION ERRORS}
\label{STATE PREPARATION ERRORS}

Here we provide the data summarized in Table \ref{StatePrepTable}.

\begin{figure}
\includegraphics[width=7.5cm]{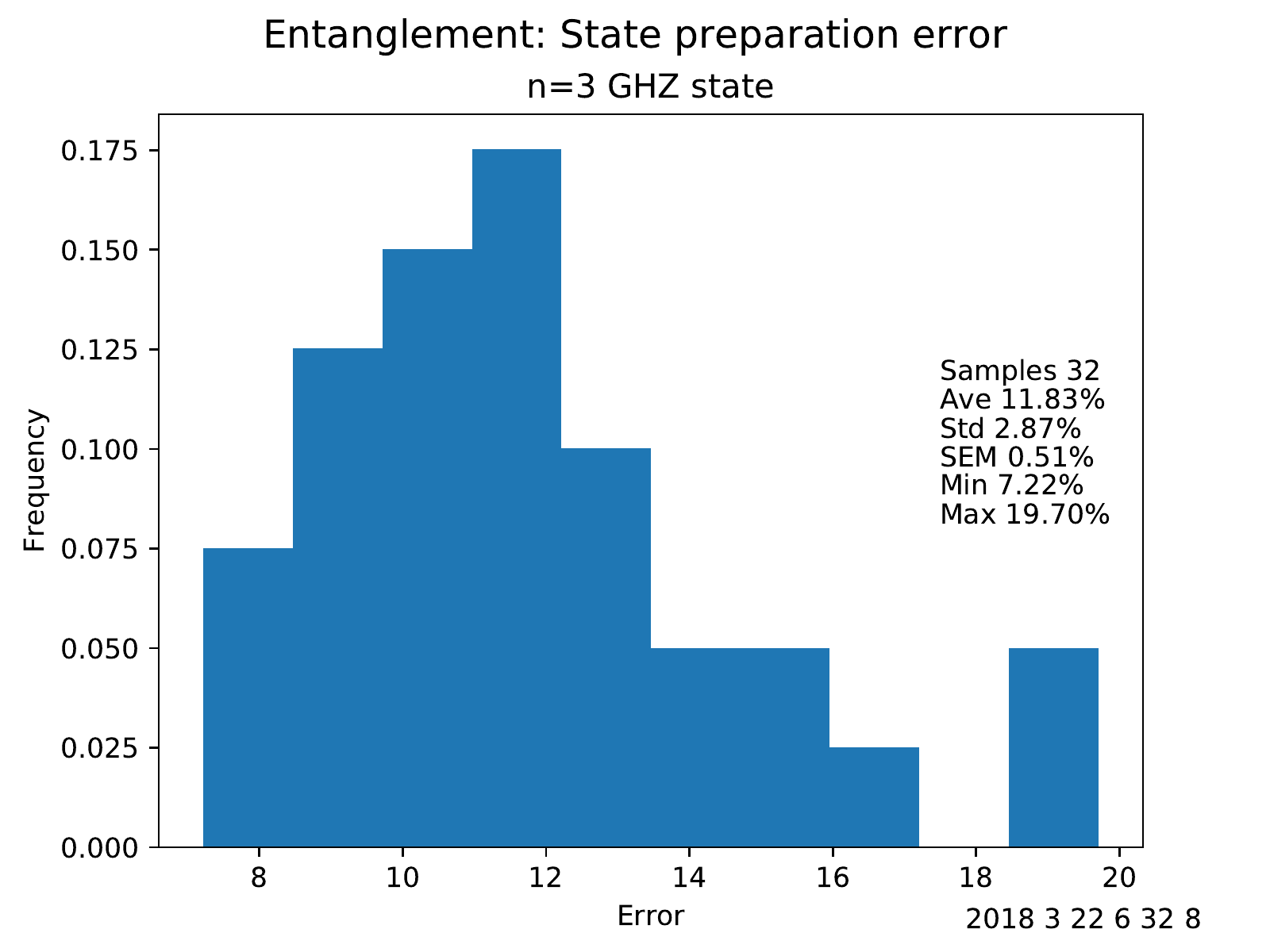} 
\caption{(Color online) Histogram of 32 independent state preparation errors for the $n\!=\!3$ GHZ state.}
\label{prepGHZ3}
\end{figure} 

\begin{figure}
\includegraphics[width=7.5cm]{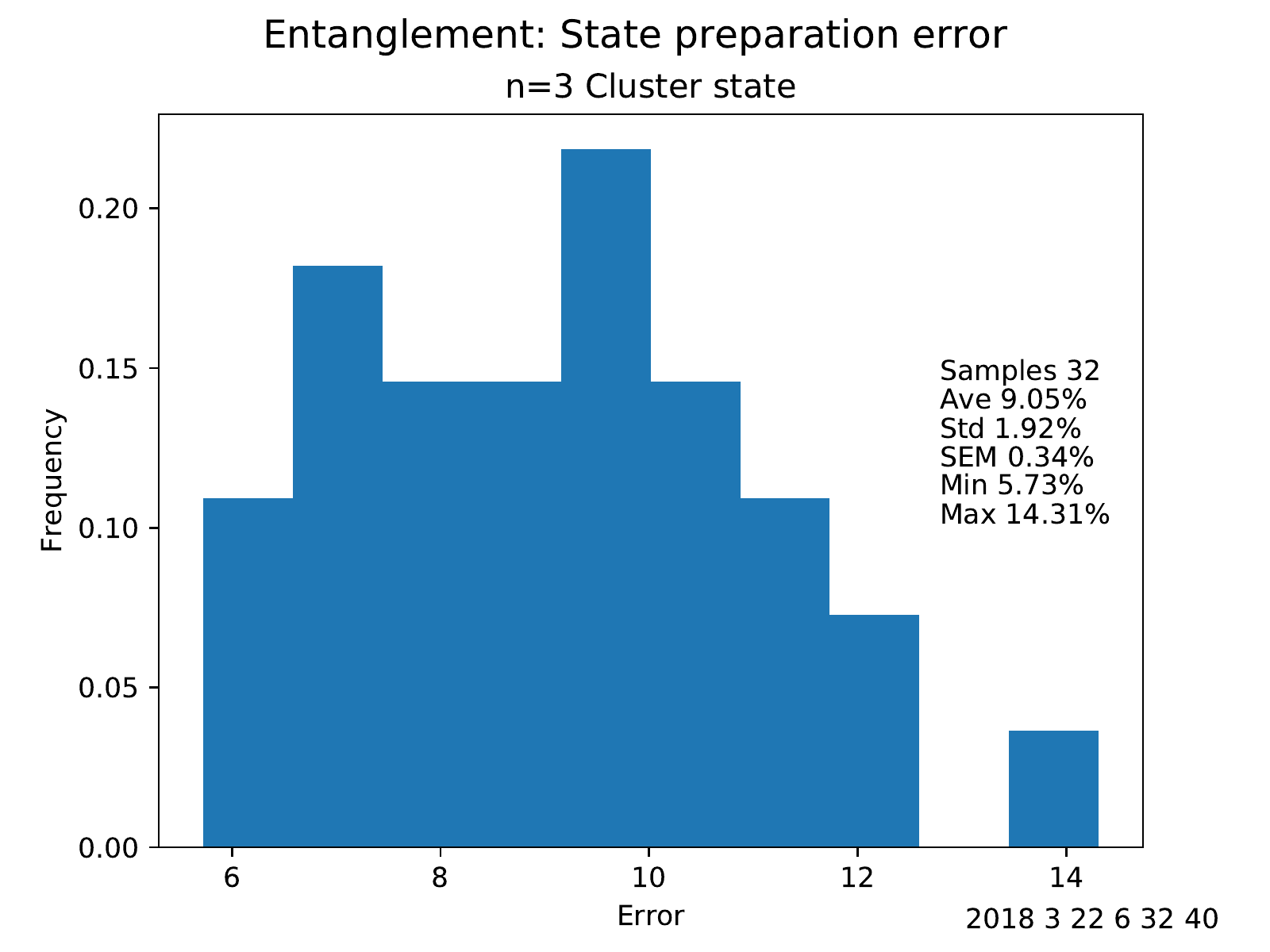} 
\caption{(Color online) Histogram of 32 independent state preparation errors for the $n\!=\!3$ cluster state.}
\label{prepGHZ3}
\end{figure} 

\begin{figure}
\includegraphics[width=7.5cm]{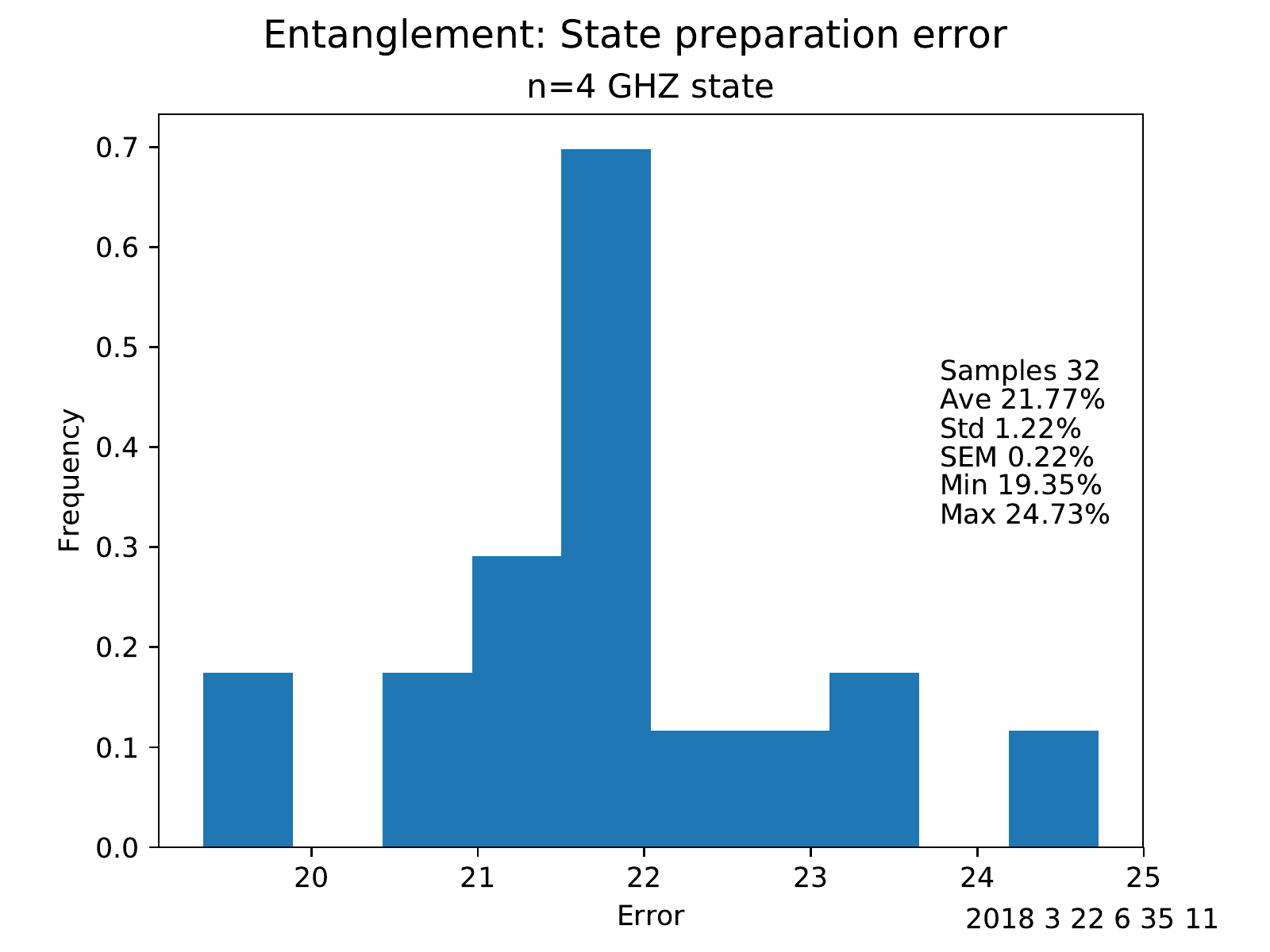} 
\caption{(Color online) Sate preparation errors for the $n\!=\!4$ GHZ state.}
\label{prepGHZ4}
\end{figure} 

\begin{figure}
\includegraphics[width=7.5cm]{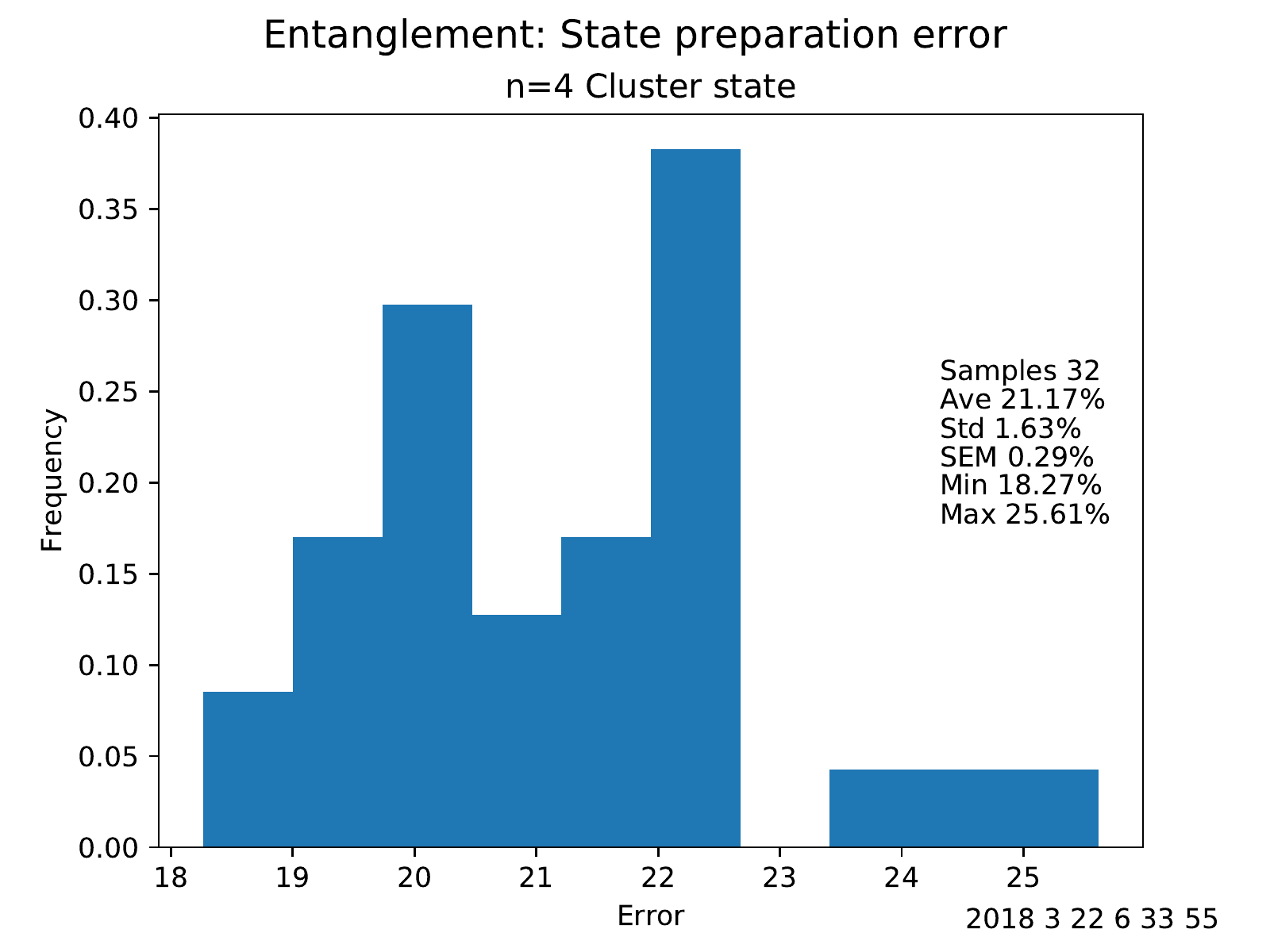} 
\caption{(Color online) State preparation errors for the $n\!=\!4$ cluster state.}
\label{prepGHZ4}
\end{figure} 

\newpage

\bibliography{/Users/mgeller/Dropbox/bibliographies/algorithms,/Users/mgeller/Dropbox/bibliographies/applications,/Users/mgeller/Dropbox/bibliographies/dwave,/Users/mgeller/Dropbox/bibliographies/control,/Users/mgeller/Dropbox/bibliographies/error_correction,/Users/mgeller/Dropbox/bibliographies/general,/Users/mgeller/Dropbox/bibliographies/group,/Users/mgeller/Dropbox/bibliographies/ions,/Users/mgeller/Dropbox/bibliographies/math,/Users/mgeller/Dropbox/bibliographies/ml,/Users/mgeller/Dropbox/bibliographies/nmr,/Users/mgeller/Dropbox/bibliographies/optics,/Users/mgeller/Dropbox/bibliographies/simulation,/Users/mgeller/Dropbox/bibliographies/superconductors,/Users/mgeller/Dropbox/bibliographies/surfacecode,endnotes}

\end{document}